\documentclass[a4paper,11pt]{article}
\pdfoutput=1 % if your are submitting a pdflatex (i.e. if you have
\usepackage{jinstpub} % for details on the use of the package, please
\usepackage[dvipsnames]{xcolor}
\usepackage{upgreek}
\usepackage{gensymb}

\makeatletter
\define@key{Gin}{myresolution}{\pdfimageresolution=#1\relax}
\makeatother

\newcommand{\comment}[1]{}

\title{Characterization of an architecture for front-end pixel binning in an integrating pixel array detector}

\author[a,b]{D. Gadkari,}
\author[a,1]{K.S. Shanks,\note{Corresponding author.}}
\author[a]{H.T. Philipp,}
\author[a]{M.W. Tate,}
\author[b]{J. Thom-Levy}
\author[a,c,d]{and S.M. Gruner}

\affiliation[a]{Laboratory of Atomic and Solid State Physics, Cornell University,\\Ithaca, NY 14853, U.S.A}
\affiliation[b]{Laboratory for Elementary-Particle Physics, Cornell University,\\Ithaca, NY 14853, U.S.A}
\affiliation[c]{Cornell High Energy Synchrotron Source (CHESS), Cornell University,\\Ithaca, NY 14853, U.S.A}
\affiliation[d]{Kavli Institute at Cornell for Nanoscale Science, Cornell University,\\Ithaca, NY 14853, U.S.A}

\emailAdd{ksg52@cornell.edu}

\abstract{Optimization of an area detector involves compromises between various parameters like frame rate, read noise, dynamic range and pixel size. We have implemented and tested a novel front-end binning design in a photon-integrating hybrid pixel array detector using the MM-PAD-2.0 pixel architecture. In this architecture, the pixels can be optionally binned in a 2$\times$2 pixel configuration using a network of switches to selectively direct the output of 4 sensor pixels to a single amplifier input. Doing this allows a trade-off between frame rate and spatial resolution. Tests show that the binned pixels perform well, but with some degradation on performance as compared to an un-binned pixel. The increased parasitic input capacitance does reduce the signal collected per x-ray as well as increases the noise of the pixel. The increase in noise is, however, less than the factor of 2 increase one would observe for binning in post-processing. Spatial scans across the binned pixels show that no measured signal intensity is lost at the inner binning unit boundaries. In the high flux regime, at a 2$\times$2 pixel wide beam spot (FWHM) size, binned mode responds linearly up to a photon flux of ~10$^{7}$ x-rays/s, and performs comparably with un-binned mode up to a photon flux of ~10$^{8}$ x-rays/s. While this study demonstrates a proof of concept for front-end binning in integrating detectors, we also identify changes to this early-stage prototype which can further improve the performance of binning pixel structures.}


\begin{document}
\maketitle
\flushbottom

\section{Introduction}
\label{sec:intro}
One of the major advantages presented by hybrid pixel array detectors (PADs) is that the separate sensor and readout layers can be optimized independently of each other. Important specifications of an area detector include frame rate, read noise, dynamic range, and pixel size. The design of an area detector inevitably involves some degree of compromise and trade-off between parameters as not all can be optimized in a single device. For example, frame rate tends to trade off with number of pixels and noise, both because more pixels require more time to read out through a given number of data channels and because electronic noise tends to increase with bandwidth. One way to cover a wider range of use cases is to make some of these trade-offs selectable, for example by allowing the user to choose to read out fewer bits in exchange for a higher frame rate at the expense of dynamic range \cite{john2012}-\cite{penni2011}, or by allowing the user to choose to bin sets of pixels together, i.e., trading spatial resolution for increased frame rate. 

In the test structures evaluated in this paper, we have implemented the latter approach in the readout ASIC (Application Specific Integrating Circuit) for a photon-integrating hybrid pixel array detector. We describe the design and characterization of an architecture for front-end binning in an integrating framework, wherein the pixel array can be optionally operated with pixels binned in a 2$\times$2 configuration for a potential factor of 4 improvement in frame rate. Because the binning architecture chosen involves additional circuitry on the pixel front end, it has potential impact on important parameters such as read noise, effective gain and integrator slew rate. 

 Previous studies have investigated pixel binning in various kinds of x-ray detectors \cite{bochenek2018}-\cite{medi5}. Pixel binning implemented before the charge-to-voltage conversion node has been common in CCDs \cite{janesick2001,howell2006}. Additionally, binning during post-processing of data using various algorithms has been well established \cite{algo1}-\cite{algo3}. Medipix3 is an example of a PAD which can be operated in ``charge summing'' mode, where signal from neighboring pixels are added together before readout \cite{medi1}-\cite{medi5} in order to deal with charge sharing between pixels. The impact of binning on read noise depends on the way binning is implemented. In typical CCD binning, where the integrated charge is summed before being converted to voltage by the readout amplifier, the read noise associated with a binned pixel unit is equal to that of an un-binned pixel. On the other hand, if binning is performed during post-processing, the read noise of each pixel within the binning unit adds in quadrature, which for a 2$\times$2 binning unit translates to an increase in noise by a factor of 2. Through our measurements, we show a significant improvement over this factor of 2 in noise enabled by our novel front-end binning architecture. In the following sections we describe this design, as well as measurements using a low-flux lab x-ray source to characterize its gain, noise and spatial response, and using a high-intensity synchrotron radiation source to characterize the high-flux performance.

\section{Front-end binning architecture}
\label{sec:arch}
%----------------------Basic pixel design------------------------
\subsection{Pixel design}
\label{subsec:det_sys}
\begin{figure}[htbp]
\centering % \begin{center}/\end{center} takes some additional vertical space
%width=.9\textwidth
%\includegraphics[myresolution=220]{MMPAD_2p0_noAG_CIN_CL_vHugh2.png}
\includegraphics[width=0.9\textwidth]{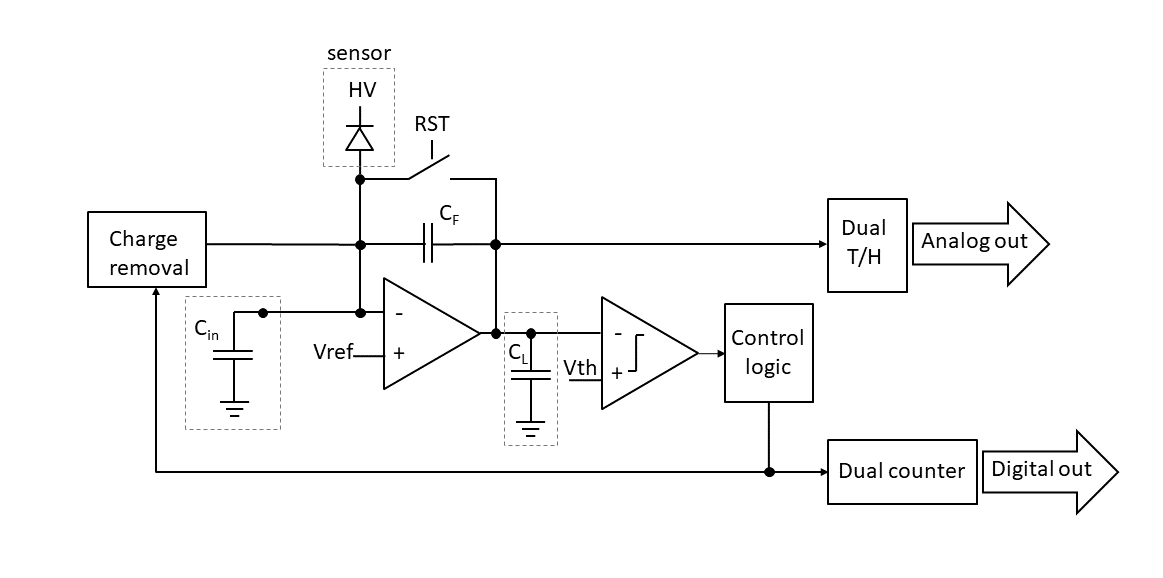}
\caption{\label{fig:mmpad2} Single-pixel MM-PAD-2.0 schematic, showing the fixed-gain variant used here. The sensor is represented by a diode on the pixel input node. The binning structures are omitted in this schematic. The front-end feedback capacitance  $C_F$ is implemented as a metal-insulator-metal (MiM) capacitor. The input capacitance  $C_{in}$ is shown for reference, and is comprised of the sensor capacitance, bond capacitance, and parasitic contributions from various CMOS components in the pixel. The load capacitance $C_L = 300$ fF is dominated by the MiM capacitors in the dual track-and-hold circuit.}
\end{figure}

Pixel binning has been implemented in a portion of a 16 $\times$ 16 pixel test chip using a version of the MM-PAD-2.0 pixel design described in \cite{weiss2017}. This pixel is a successor of the original Mixed Mode Pixel Array Detector (MM-PAD) developed at Cornell \cite{schuette2008,tate2013}. While \cite{weiss2017} described an MM-PAD-2.0 variant with adaptive front-end gain, the pixel variant used here implemented a fixed front-end gain for the sake of simplicity. A single-pixel schematic is shown in figure \ref{fig:mmpad2}.

% MM-PAD-2.0 ASIC diagram without adaptive capacitance. 

The MM-PAD family of pixels all use a photon-integrating pixel front-end coupled with a charge removal mechanism to achieve an extended dynamic range. A relatively small integration capacitor (in the MM-PAD-2.0 pixel, 40 fF) is used to maintain reasonable single-photon resolution in the low-signal regime. When the output of the front-end integrator reaches an externally-set threshold $V_{th}$, a charge removal circuit is triggered and removes a fixed amount of charge from the integration capacitor, without interrupting integration of additional incoming photocurrent. In the MM-PAD-2.0 pixel, charge equivalent to about 400 keV total deposited energy is removed per charge removal operation. An in-pixel counter keeps track of the number of charge removals in a given exposure. At the end of an exposure, the counter value as well as the residual analog voltage at the integrator output are both read out. In a fully-calibrated system, the digital and analog portions are scaled together to reconstruct the total signal detected during the exposure. Dual track-and-hold circuits and dual in-pixel counters are used to achieve readout with approximately 98\% duty cycle \cite{hugh2019}.
%----------------------Binning implementation------------------------
\begin{figure}[htbp]
\centering 
%width=.45\textwidth
%\includegraphics[myresolution=360]{binning_schematic2.png}
\includegraphics[width=.75\textwidth]{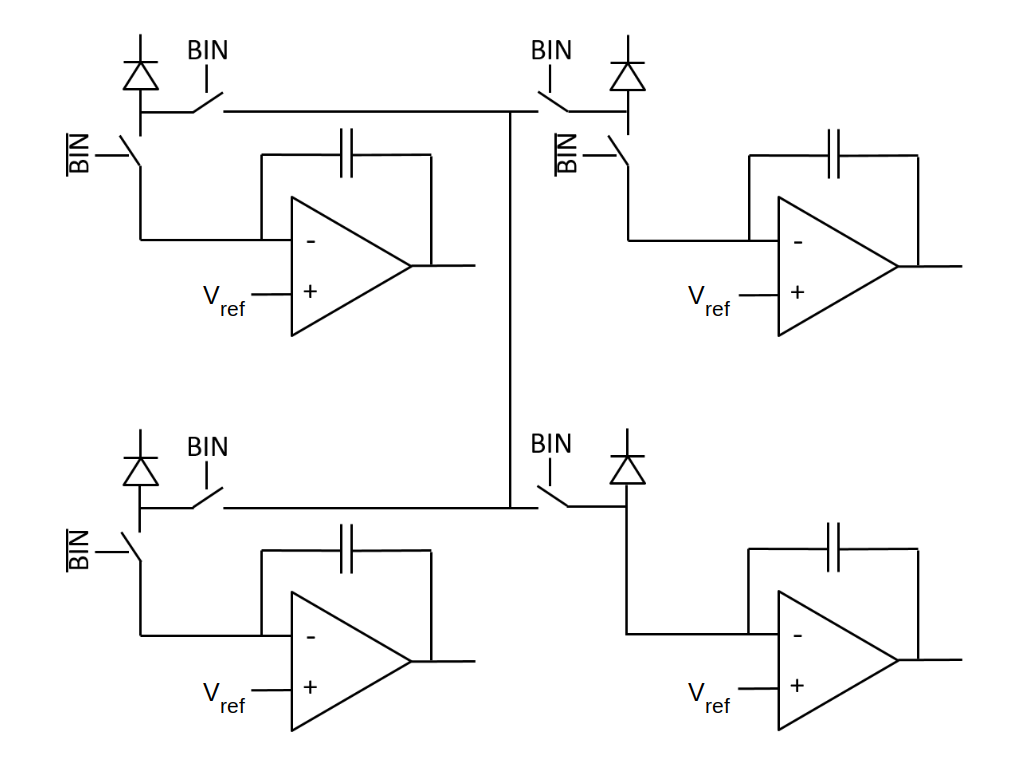}
\caption{\label{fig:schem} Schematic of a four-pixel binning network. The four amplifiers shown are the input amplifiers of figure \ref{fig:mmpad2} for 4 adjacent pixels. For simplicity, only the front-end integrator of each pixel is shown, with the pixel reset switch (shown in figure 1) omitted. When binned, all four pixels act as a “super-pixel” that is read out through a “master pixel” amplifier (lower right).} 
\end{figure}

Figure \ref{fig:schem} shows the four-pixel binning scheme. In general, the input capacitance of the pixel, $C_{in}$, depicted in figure \ref{fig:mmpad2}, has contributions from CMOS parasitics on the ASIC, the capacitance of the pixel bond pad to the ASIC substrate, the capacitance of the sensor, and contributions from the bond itself. Compared to a pixel with no binning structures, the binning switches themselves add a small additional parasitic capacitance which is present whether binning is on or off. Also, when binning is enabled the master pixel sees additional capacitance from each of the other three pixels in the 2$\times$2 binning unit, consisting of the sensor and bond capacitances plus any CMOS parasitics that are upstream of the binning switches.  

Increasing $C_{in}$ has the effect of decreasing the apparent gain of the pixel front-end by reducing the charge collection efficiency of the amplifier. Incoming photogenerated charge from the sensor is not purely integrated onto the feedback capacitor $C_F$ but instead split between $C_F$ and $C_{in}$. The charge collection efficiency is given by

\begin{equation}
\label{eq:cce}
CCE = \frac{(1+A)C_F}{C_{in}+(1+A)C_F}
\end{equation}

\noindent where $A$ is the open-loop voltage gain of the amplifier. This can be viewed as an equivalent decrease in gain assuming the ideal case of zero $C_{in}$ and an effective feedback capacitance of $C_{F,eff}$ given by

\begin{equation}
\label{eq:cfeff}
C_{F,eff} = C_F\frac{C_{in}+(1+A)C_F}{(1+A)C_F}.
\end{equation} 

\noindent In the pixel under consideration here, $A$ is approximately 40 and $C_F$ is 40 fF. $C_{in}$ is empirically determined to be 484 fF for un-binned mode and 767 fF for binned mode, as shown in section \ref{subsec:gain}.

%previous language:
%An increase in $C_{in}$ when binning is switched on is also expected to increase the pixel noise. A change in $C_{in}$ will impact the $kTC$ noise from the input capacitance at the start of an exposure, and the transfer function governing the amplifier read noise. 
%new language 7/17/2020:
An increase in $C_{in}$ when binning is switched on is also expected to increase the noise seen at the pixel output node. This noise has contributions from the amplifier itself (\textit{i.e.}, transistor thermal and flicker noise) and from switching noise associated with the opening of the pixel reset switch (shown in figure \ref{fig:mmpad2}) at the start of an exposure. A quantitative estimation of these relative noise contributions is beyond the scope of this paper; detailed considerations of the impact of $C_{in}$ on amplifier noise and switching noise can be found in \cite{schreier2005} and \cite{kozlowski2005, spieler2005}, respectively. A brief qualitative discussion of the expected dependence on $C_{in}$ follows.

We will refer to the combination of these noise sources, and any others in the analog readout chain, as the pixel read noise. In the context of an x-ray detector, it is useful to consider the read noise in units of equivalent noise charge, since this allows for quick comparison with the signal produced by an individual x-ray stopped in the sensor, which is given in units of charge by the ratio of the x-ray energy to the energy required to produce an electron-hole pair in the sensor (3.65 eV in the case of silicon).

The integration of charge $Q$ onto $C_{F,eff}$ produces a change in voltage $V=Q/C_{F,eff}$ at the pixel output. As such, the output-referred amplifier voltage noise $V_{n,amp}$ can be interpreted as an equivalent noise charge $Q_{n,amp} = C_{F,eff}V_{n,amp}$. $V_{n,amp}$ is given by

\begin{equation}
\label{eq:amp_noise}
V_{n,amp}^2 = S\frac{G_m(C_F+C_{in})^2}{(C_F+C_{in})C_FC_L+C_F^2C_{in}}
\end{equation}

\noindent where $S$ is the power spectral density of the amplifier noise as seen as a voltage source at the non-inverting input and $G_m$ is the amplifier transconductance. $S$ and $G_m$ are properties of the amplifier itself, and are unchanged by the state of the binning switches. Since $C_F$ and $C_L$ are also fixed, an increase in $C_{in}$ will lead to an increase in $V_{n,amp}$. Furthermore, from equation \ref{eq:cfeff}, the effective feedback capacitance $C_{F,eff}$ also increases with $C_{in}$. These factors together lead to an increase in $Q_{n,amp}$, the amplifier's contribution to the total pixel read noise. 

The switching noise associated with the opening of the pixel reset switch at the beginning of an exposure, also referred to as the pixel's $kTC$ noise, can be understood as arising from sampling, at the moment the reset switch opens, the thermal fluctuation of charge on the capacitors connected to the pixel input node. This gives a contribution from $C_{in}$ to the pixel read noise, in units of equivalent noise charge, that scales as $\sqrt{kTC_{in}}$ \cite{kozlowski2005,spieler2005,koerner2010}. Because the binning switches (shown in figure \ref{fig:schem}) are configured to be either open or closed while the pixel is in reset before the start of an exposure, and do not switch states during an exposure, they do not contribute switching noise to the total pixel read noise.

A third contribution to the read noise arises from the injection of charge from the reset switch onto the pixel input node when the switch is opened at the start of an exposure \cite{wegmann1987}. This results in a noise charge contribution that depends on the switch size and is expected to be independent of $C_{in}$. However, because the reset switches in the pixel examined here are large, this contribution to the read noise is significant and may be on par with the amplifier noise charge $Q_{n,amp}$. Circuit simulations suggest that these two noise sources dominate over the $kTC$ noise in the pixel examined here.

\subsection{Small-scale detector system}
\label{subsec:ss_det_sys}

A small-scale detector module was assembled consisting of a 16$\times$16 pixel readout ASIC stud bonded (Polymer Assembly Technology, Michigan, USA) to a 500 $\upmu$m thick Si sensor (SINTEF, Oslo, Norway) with a pixel pitch of 150 $\upmu$m $\times$ 150 $\upmu$m. The detector module was wire-bonded into a ceramic pin grid array package, which can be plugged into a zero-insertion force socket on a host PCB that supplies power, current and voltage biases. A Virtex 6 FPGA resident on an Xilinx ML605 development board handles detector control waveform generation and data capture. Captured data is assembled into frames and transmitted to a host computer over Ethernet for storage and analysis. 

\begin{figure}[htbp]
\centering 
\includegraphics[myresolution=220]{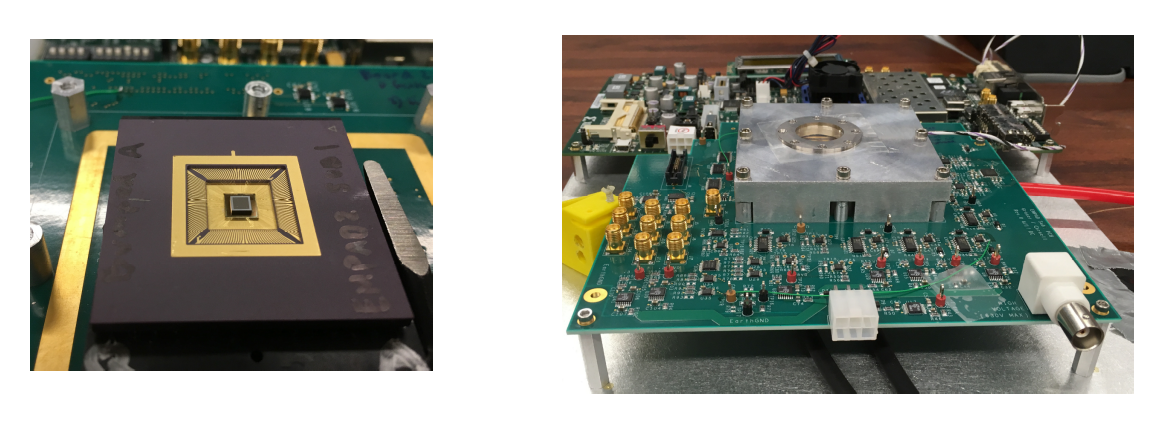}
\caption{\label{fig:house_fpga}(Left) Wire bonded chip with attached sensor in a pin grid array package. (Right) ``Clamshell'' vacuum housing mounted around the detector socket, to provide a condensation free environment for testing the system at low temperatures. A bottom portion of the clamshell (below the printed circuit board and not seen in the figure) houses a thermoelectric module used to cool the detector.} 
\end{figure}

During operation, the detector module temperature is held at -20\degree C via a thermoelectric cooler. To prevent water condensation, vacuum is maintained around the detector via a ``clamshell'' housing with an x-ray transparent window fitted around the detector module. Figure \ref{fig:house_fpga} shows the wire-bonded detector inserted in the socket (left), and the vacuum housing with the host PCB and ML605 board (right).

Within the ASIC, the pixels are organized into four banks of 4$\times$16 pixels each. Each bank is a variant of the MM-PAD-2.0 pixel and each has its own analog and digital readout channel. Binning is implemented only on one bank (``bank 1'' in this report). A neighboring bank (``bank 2'') has an identical pixel design with no binning structures. Both of these banks have a 40 fF feedback capacitor with no adaptive gain.

% detector system photo

\section{Low-flux characterization}
The gain and noise of the detector in binned mode and un-binned mode were measured and compared. The spatial response of the 2$\times$2 binning unit was also tested by raster scanning x-rays across the detector using a tungsten pinhole mask. 

\subsection{Signal measurement and analysis method}
The signal from a MM-PAD-2.0 pixel consists of a digital number corresponding to the number of charge removal steps, plus the analog value remaining on the output of the charge integrator at the end of the frame. The analog values are digitized to 14 bits and are scaled to and combined with the digital counts to produce a wide-dynamic-range intensity value, here reported in Analog-to-Digital units (ADU) of the analog measurements. The output is offset corrected by subtracting the average of multiple dark frames taken with no x-ray signal \cite{becker2018}.

\subsection{Pixel gain measurements}
\label{subsec:gain}
An x-ray spectral histogram can be used to determine the gain of each pixel in binned and un-binned mode. Here, a 150 $\upmu$m thick tungsten pinhole mask with 25 $\upmu$m holes is used to isolate the illuminated region to the central portion of a pixel. Integration time was chosen such that the signal would correspond to only a few x-rays/pixel/frame. The mask was translated to provide datasets for each pixel. X-rays were produced via a 50 W silver anode tube operated at 45 kV and 0.4 mA. A graphite monochromator was used to isolate the 22.16 keV Ag K-alpha line.

\begin{figure}
\centering 
\includegraphics[myresolution = 300]{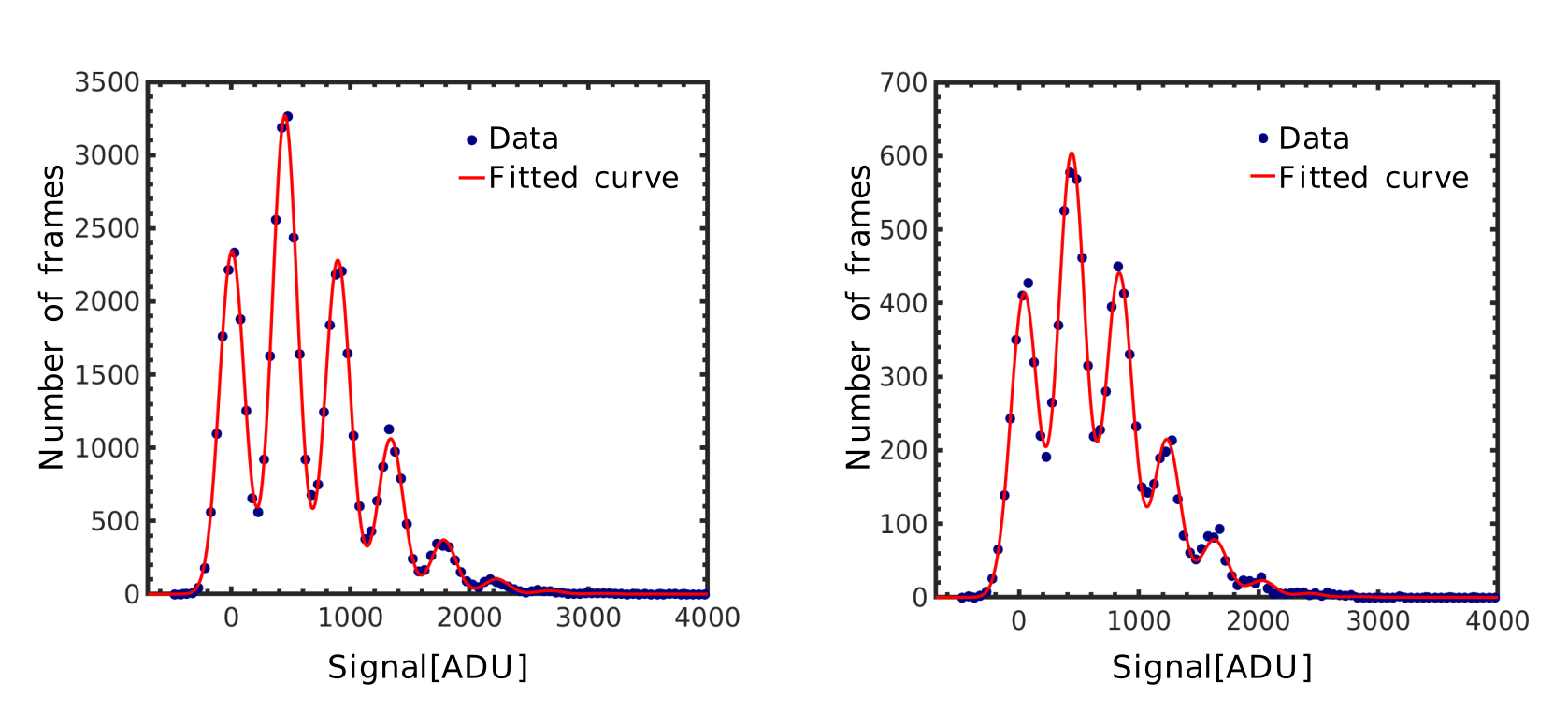}
\caption{\label{fig:pinhole_spectra} Pinhole spectra for pixel on row 4, column 7 when (left) binning is switched off, and (right) binning is switched on. The average gain for the pixels is measured to be 433 ADU/22.16 keV in un-binned mode and 382 ADU/22.16 keV for binned mode. The gain is reduced by by 12\% in the binned mode as compared to the un-binned mode.}
\end{figure}

	For each pixel, 10,000 frames for binned mode and 50,000 frames for un-binned mode  were captured, each with integration time of 12.5 ms. To determine the gain of a pixel, the signal recorded by that pixel is histogrammed across all frames; the result is a series of peaks corresponding to integer numbers of photons (i.e. 0, 1, 2, ...) detected by that pixel during the integration window, as shown in figure \ref{fig:pinhole_spectra}. The distance between consecutive peaks gives the gain (in ADU/ph) of the pixel for a single photon. 

	  A sum of Gaussians was fit to the data to extract the gain. The height of each Gaussian is constrained to follow a Poisson distribution, and the width of each Gaussian is constrained to be equal. The following equation shows the fit function for a histogram containing $n$ photon peaks, where $G$, $\sigma$ and $h$ are free parameters to be fit:

\begin{equation*} \label{eq:fit_func}
 f(x)  =  \sum_{k=0}^{n}  \frac{h}{k!}\left(\frac{x_{av}}{G}\right)^k \exp\left(-\frac{x_{av}}{G} \right) \exp\left( -\left(\frac{x-kG}{\sigma}\right)^{2} \right)
\end{equation*}

\noindent where $x$ denotes the signal in the pixel, $G$ denotes the gain of the pixel, $\sigma$ denotes the width of each Gaussian photon peak, and $x_{av}$ is the average signal falling on the pixel across all frames. Figure \ref{fig:pinhole_spectra} (left) shows the x-ray spectrum for pixel on row 4 and column 7 in un-binned mode. Figure \ref{fig:pinhole_spectra} (right) shows the x-ray spectrum for the same pixel for binned mode. It is important to note here that for binned mode, the measured gain in any pixel refers to when the photons are incident on that pixel but are still read out through master pixel of the corresponding master pixel of the 2$\times$2 unit. The average gain measured for the detector in un-binned mode is 433$\pm$9 ADU/22.16 keV and binned mode is 382$\pm$24 ADU/22.16 keV. Hence, the gain decreases by 12$\pm$6\% when binning is switched on. 

From the measured gain we obtain the effective feedback capacitance, $C_{F,eff}$, for un-binned and binned mode, which is calculated to be 52 fF and 59 fF for the two modes respectively. Using equation \ref{eq:cfeff}, we calculate the total input capacitance, $C_{in}$, of the pixel to be 484 fF for un-binned mode and 767 fF for binned mode. From this we estimate a capacitance of 94 fF per pixel upstream of the binning switches, with 390 fF per pixel downstream. The downstream capacitance is due, in large part, to various test structures included in this chip that are unrelated to pixel binning. A significant reduction in $C_{in}$ can be expected for future ASICs.

% Pixel spectrum 

\subsection{Pixel noise measurements}
System noise was measured with and without binning enabled by comparing 6700 frames with a short integration time (20 ms) with no added signal. The dark charge accumulated during this short period added neglible noise to the overall measurement. 
	 
	In un-binned mode, the average read noise of the binning-capable pixels was 107$\pm$19 ADU rms (5.48$\pm$0.89 keV). No difference is seen between the master binning pixels and the secondary pixels. This was the same noise seen for pixels in bank 2 which have no binning structures. When binning was turned on in bank 1, the master pixels had an average noise of 175$\pm$28 ADU (10.15$\pm$1.74 keV). Hence, the noise goes up by 63$\pm$8\% when binning is switched on. An increase in noise is expected due to the increased input capacitance in the binned case. We note that the noise in the now unconnected secondary pixels drops to 92$\pm$16 ADU. This is due to the reduction in input capacitance in this case. 

\subsection{Spatial response measurements}
 
\begin{figure}
\centering 
\includegraphics[width=.99\textwidth]{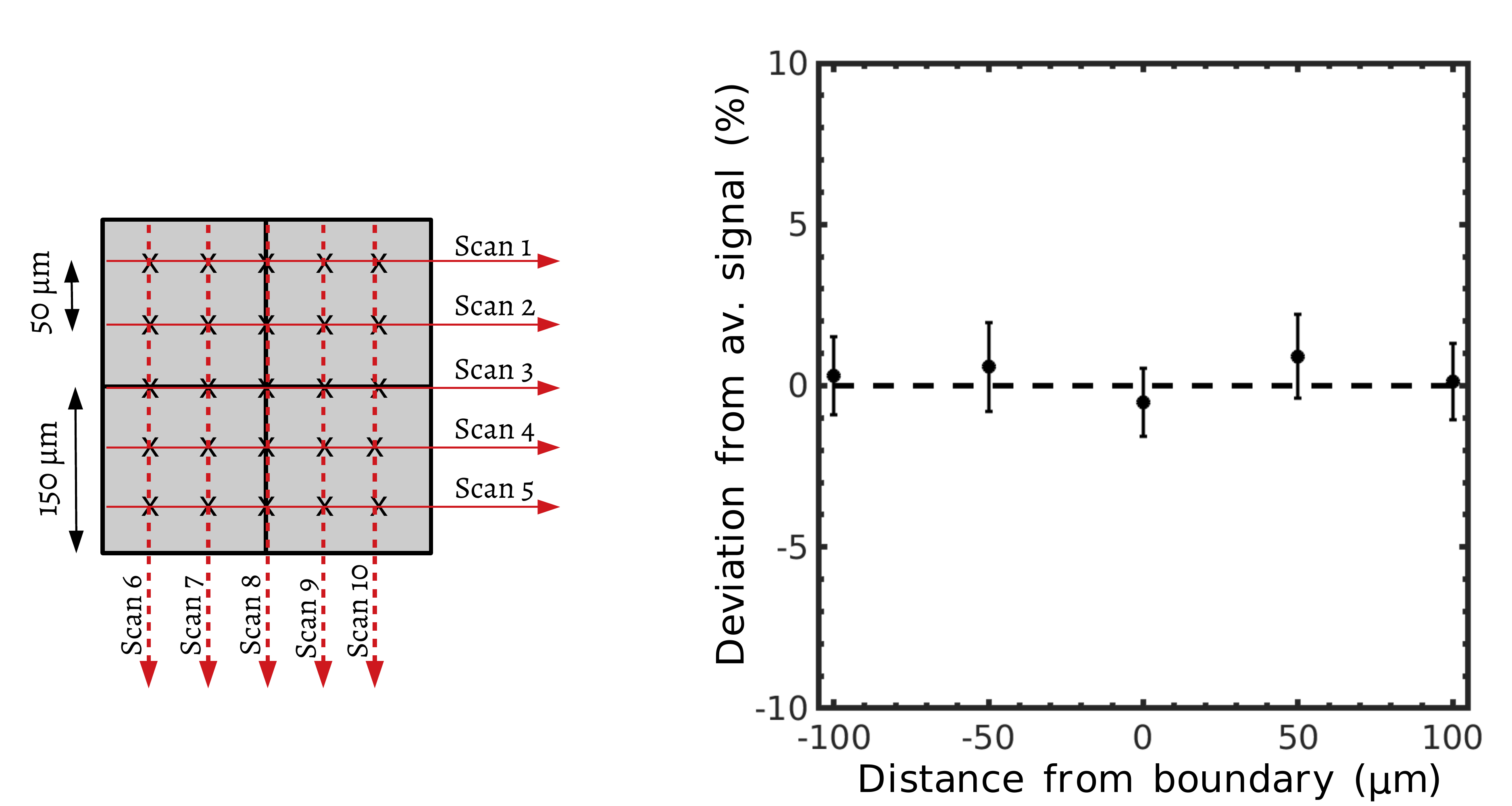}
\caption{\label{fig:raster_cartoon} (Left) Cartoon depiction of the positions at which signal intensity was measured for a binning unit in binned mode, depicted by `x' marks. The spatial response of the detector in binned mode is obtained in terms of the deviation of measured signal intensity, in horizontal and vertical direction, and averaged together. (Right) Percent deviation of the signal averaged together for all the raster scans, versus the distance of pinhole from the inner pixel boundaries. The response of the super-pixel unit is constant and varies less than 1\%, with no loss at the inner pixel boundaries.}
\end{figure}

 To confirm that no signal is lost in binning mode, especially at pixel boundaries within a 2$\times$2 binning unit, a raster scan was conducted across bank 1. A 25 $\upmu$m pinhole was shifted across each binning unit in steps of 50 $\upmu$m in both the vertical and horizontal directions to get 25 different measurements of signal intensity, as shown in figure \ref{fig:raster_cartoon} (left). A 50 W silver anode tube was operated at 45 kV and 0.4 mA, with a graphite monochromator to isolate x-rays of 22.16 keV.
 
 For each pinhole position, a total of 10,000 frames were captured with an integration time of 20 ms. The deviation of the measured signal as a function of distance from the inner binning unit boundary was averaged across different binning units. The variation in the signal is measured to be less than 1\% across the unit, as shown in figure \ref{fig:raster_cartoon} (right). The uncertainty in the signal deviation is attributed to the variation in the photon flux from the x-ray source. No loss of signal is observed at the boundaries within a binning unit. The gain was also measured across the same grid points using the method descibed above. Measurements of gain at the interior pixel boundaries of the binning unit using the spectral method agreed with measurements in the pixel interior within 1.6\%. Additionally, no difference was seen between any of the four sub-pixels in the binning unit.

% Raster scan result and cartoon  

\section{High-flux characterization}

High-flux measurements were conducted at beamline 4B of the Cornell High Energy Synchrotron Source (CHESS), which recently underwent an upgrade where a double-bend achromat lattice is used to achieve high flux density \cite{shanks2019}. The beamline is fed by a 1.5m long CHESS Compact Undulator (CCU) \cite{temnykh2015}-\cite{temnykh2013}. A diamond <111> monochromator was used to set the x-ray energy to 11.2 keV. Slits were used to set the beam size to 100 $\upmu$m$\times$100 $\upmu$m approximately 1 m upstream of the detector. Due to beam divergence, the spot size on the detector was about 2$\times$2 pixels (FWHM) wide. An ionization chamber immediately downstream of the slits provided an independent measure of the incident flux. Throughout the measurement, the direct beam intensity as measured by the ionization chamber was stable to within 5\%. A set of seven 10 $\upmu$m thick copper foils was placed directly in front of the detector and removed one at a time in order to vary the flux incident on the detector from 10$^{6}$ x-rays/s to 10$^{10}$ x-rays/s. 

At each attenuation step, the ionization chamber reading was averaged over 20 s, and a series of 500 exposures, each of 200 $\upmu$s, was acquired.  The flux incident on the detector at each step was computed from the average ionization chamber reading and the known attenuation of the Cu foils. A correction factor for the 800 mm of air after the ionization chamber was applied. The detector measurement integrated the entire beam spot over the size of 2$\times$2 pixels (FWHM).

\begin{figure}
\centering 
\includegraphics[width=.7\textwidth]{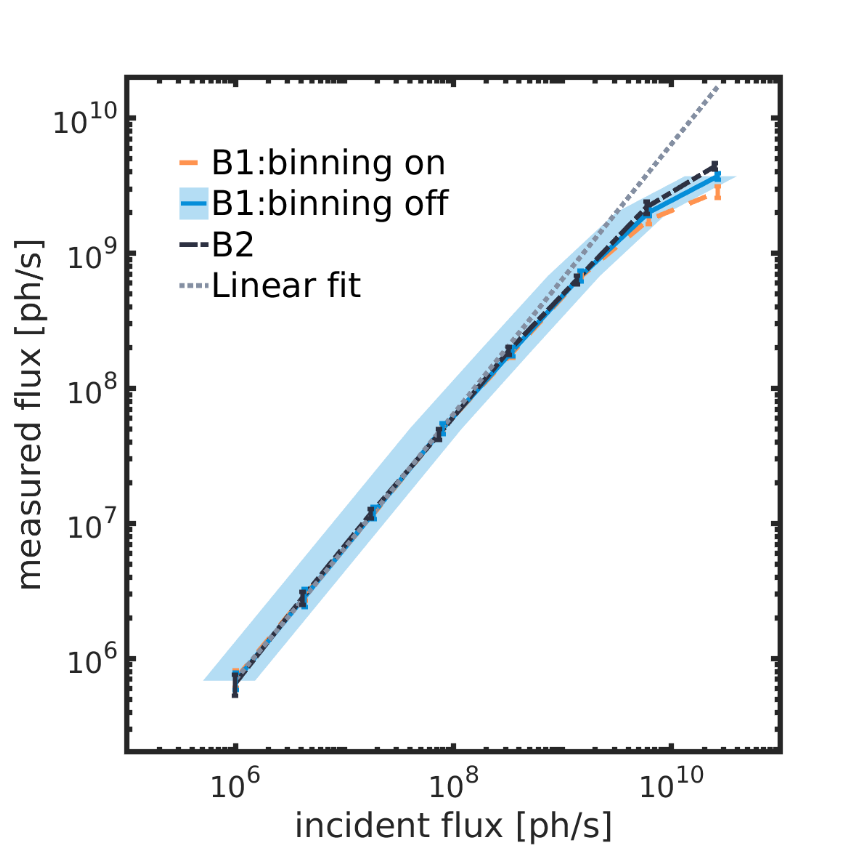} %inclusive_color_band
\caption{\label{fig:high_flux} The measured vs incident flux (x-rays/s) in the first and the second bank. In the first bank, the response of binned and un-binned mode were measured. A linear fit for the initial region of the curves is displayed. The blue band depicts the uncertainty on the incident flux, dominated by the systematic uncertainty in the ionization chamber gain factor, shown here only for the blue curve. The other curves have the same uncertainty associated with the incident flux. Binned mode performs comparably with the other two curves in the linear region. The binned curve deviates from the un-binned curve in the high flux region, since the flux incident on 4 pixels is now handled by only one front-end amplifier.}
\end{figure}

 Figure \ref{fig:high_flux} shows the flux measured by the detector versus the incident flux measured by the ionization chamber. Included are curves for bank 1 in binned and un-binned mode, as well as bank 2. The largest uncertainty is a systematic uncertainty in the gain factor of the ionization chamber readings. All three cases have a linear response up to a flux of at least 4$\times$10$^{7}$ x-rays/s. For context, most photon counting PADs see a count rate limit of $10^6-10^7$ s$^{-1}$ \cite{trueb2012}. Behavior of the binned pixel matches that of the un-binned pixel up to at least 7$\times$10$^{8}$ x-rays/s. At higher rates, the binned pixel performance begins to roll off from that of the un-binned pixel. This is expected behavior due to the fact that the flux incident over 4 pixels is processed by only one front-end amplifier when binning is enabled. The additional input capacitance when binning is enabled does not appear to have an additional adverse affect on the high-flux performance. As noted in section \ref{subsec:ss_det_sys}, adaptive front-end gain was not implemented in either banks 1 or 2. Combining adaptive gain with binning would further improve the high-flux performance of the pixel in terms of maximum sustained flux capability.
 
 % high flux plot
 
\section{Conclusions} 

%This test ASIC had many pixel structures being evaluated concurrently, and as such, the pixel was not optimized for noise performance.

We have implemented a novel front-end binning architecture in a photon-integrating PAD. The gain of the pixel decreased and the noise increased, as anticipated, due to the increased input capacitance seen when 4 pixels are binned together. The increase in noise is nevertheless less than the increase in noise one would obtain via 2$\times$2 pixel binning in post-processing. Measurements of the spatial response showed no loss of signal at pixel boundaries within a binning unit.

This ASIC was an early-stage prototype intended, in part, to provide a proof of concept of the front-end binning architecture. No attempt was made to implement a readout architecture for the 16$\times$16 pixel array (i.e., pixel addressing and multiplexing during readout) that could realize the increased frame rate that is possible via 2$\times$2 pixel binning. Any future binning implementation would be accompanied by a streamlined readout architecture capable of realizing the approximately 4$\times$ improvement in frame rate. 

Improvements in overall performance, particularly in terms of noise and input capacitance, are possible. We note that a number of structures were included in this test chip, such as input protection diodes which will have an inpact on the total input capacitance. One contribution to $C_{in}$ that deserves particular attention is any overlap between the bump bond and any underlying metallization on the ASIC, e.g., power planes, which often cover most of the pixel area. The ASIC passivation is often thin with a relatively high dielectric constant, and therefore has the potential to contribute significantly to $C_{in}$. This can be mitigated by adding an additional, thicker passivation layer during ASIC post-processing. We also note that the pixel reset switches, as implemented in this pixel, were quite large to evaluate behavior at extremely high fluxes. These switches contribute significantly to the overall input capacitance. Through careful design, we expect that a significant reduction in input capacitance can be made in future devices, leading to both improved CCE and a reduction in noise. High-flux performance can also be expected to improve if an adaptive gain architecture is added to the front end.

\acknowledgments
The authors would like to thank Jacob Ruff at Cornell High Energy Synchrotron Source (CHESS) and Hannah Hu at Cornell University for their help with the high-flux experiments at CHESS. This work is based upon research conducted at the Center for High Energy X-ray Sciences (CHEXS) which is supported by the National Science Foundation under award DMR-1829070. Detector development at Cornell is supported by Thermo Fisher Scientific, U.S. Department of Energy grant DE-SC0017631 and the Cornell High Energy Synchrotron Source (CHESS) under U.S. National Science Foundation Award DMR-1332208.



\begin{thebibliography}{99}

\bibitem{john2012}
I. Johnson, A. Bergamaschi, J. Buitenhuis, R. Dinapoli, D. Greiffenberg, B. Henrich, T. Ikonen, G. Meier, A. Menzel, A. Mozzanica, V. Radicci, D. K. Satapathy, B. Schmitt and X. Shi
\emph{Capturing dynamics with Eiger, a fast-framing X-ray detector.}, 
\emph{J. Synchrotron Rad. } {\bf 19} (2012) 1001-1005.

\bibitem{penni2011}
 D. Pennicard, S. Lange, S. Smoljanin, J. Becker, H. Hirsemann, M. Epple and H. Graafsma 
\emph{Development of LAMBDA: Large Area Medipix-Based Detector Array.}, 
\emph{JINST} {\bf 6} (2011) C11009 .

\bibitem{bochenek2018}
M. Bochenek, S. Bottinelli, Ch. Broennimann, P. Livi, T. Loeliger, V. Radicci, R. Schnyder, and P. Zambon
\emph{IBEX: Versatile Readout ASIC With SpectralImaging Capability and High Count Rate Capability.}, 
\emph{IEEE  Trans. Nucl. Sci.} {\bf 65(6)} (2018) 1285-1291 .

\bibitem{janesick2001}
J. R. Janesick.
\emph{Scientific Charge-coupled Devices}, 
SPIE Press, Bellingham Washington U.S.A. (2001).

\bibitem{howell2006}
S. B. Howell
\emph{Handbook of CCD Astronomy}, in \emph{Cambridge Observing Handbooks for Research Astronomers} \textbf{5}, 
Cambridge University Press (2006).

\bibitem{algo1}
J.S. Sanders and A.C. Fabian 
\emph{Adaptive binning of X-ray galaxy cluster images.}, 
\emph{Monthly Notices of the Royal Astronomical Society} {\bf 325(1)} (2001) 178-186.

\bibitem{algo2}
M. Cappellari and Y. Copin 
\emph{Adaptive spatial binning of integral-field spectroscopic data using Voronoi tessellations.}, 
\emph{Monthly Notices of the Royal Astronomical Society} {\bf 342(2)} (2003) 345-354.

\bibitem{algo3}
Y. Yoo, J. Im and J. Paik 
\emph{Low-Light Image Enhancement Using Adaptive Digital Pixel Binning.}, 
\emph{Sensors} {\bf 15(7)} (2015) 14917-14931.


\bibitem{medi1}
R. Ballabriga, M. Campbell, E. Heijne, X. Llopart, L. Tlustos, W. Wong 
\emph{Medipix3: A 64 k pixel detector readout chip working in single photon counting mode with improved spectrometric performance.}, 
\emph{Nuclear Instruments and Methods in Physics Research A} {\bf 633} (2011) S15-S18.

\bibitem{medi2}
D. Pennicard, R. Ballabriga, X. Llopart, M. Campbell, H. Graafsma 
\emph{Simulations of charge summing and threshold dispersion effects in Medipix3}, 
\emph{Nuclear Instruments and Methods in Physics Research A} {\bf 636} (2011) 74-81.

\bibitem{medi3}
R. Ballabriga, M. Campbell, E. H. M. Heijne, X. Llopart, L. Tlustos
\emph{The Medipix3 Prototype, a Pixel Readout Chip Working in Single Photon Counting Mode With Improved Spectrometric Performance}, 
\emph{IEEE Trans. Nucl. Sci.} {\bf 54} (2007) 1824-1829.

\bibitem{medi4}
E. N. Gimenez, R. Ballabriga, M. Campbell, I. Horswell, X. Llopart, J. Marchal, K. J. S. Sawhney, N. Tartoni, D. Turecek
\emph{Characterization of Medipix3 With Synchrotron Radiation}, 
\emph{IEEE Trans. Nucl. Sci.} {\bf 58} (2011) 323-332.

\bibitem{medi5}
R. Ballabriga, G. Blaj, M. Campbell, M. Fiederle, D. Greiffenberg, E.H.M. Heijne, X. Llopart, R. Plackett, S. Procz, L. Tlustos, D. Turecek, W. Wong.
\emph{Characterization of the Medipix3 pixel readout chip}, 
\emph{JINST} {\bf 6} (2011) C01052.

\bibitem{weiss2017}
J. T. Weiss, K. S. Shanks, H. T. Philipp, J. Becker, D. Chamberlain, P. Purohit, M. W. Tate, S. M. Gruner. \emph{High Dynamic Range X-ray Detector Pixel Architectures Utilizing Charge Removal.}, 
\emph{IEEE Trans. Nucl. Sci.} {\bf 64} (2017) 1101-1107.

\bibitem{schuette2008}
D. R. Schuette.
\emph{A mixed analog and digital pixel array detector for synchrotron x-ray imaging.}, 
\emph{Ph.D. dissertation} (2008) Cornell University, USA.
 
\bibitem{tate2013}
M.W. Tate, D. Chamberlain, K.S. Green, H.T. Philipp, P. Purohit, C. Strohman, S.M. Gruner.
\emph{A medium-format, mixed-mode pixel array detector for kilohertz x-ray imaging.}, 
\emph{J. Phys.} Conf. Ser. 425 062004 (2013).

\bibitem{hugh2019}
H.T.Philipp, M.W.Tate, K.S.Shanks, P.Purohit, S.M.Gruner.
\emph{Practical considerations for high-speed X-ray pixel array detectors and X-ray sensing materials.}, 
\emph{Nuclear Instruments and Methods in Physics Research A} {\bf 925} (2019) 18-23.

\bibitem{schreier2005}
R. Schreier, J. Silva, J. Steensgaard, G. C. Temes
\emph{Design-oriented esti-mation of thermal noise in switched-capacitor circuits}, 
\emph{IEEE Transactions on Circuits and Systems - I} {\bf 52} (2005) 2358–2368.

\bibitem{kozlowski2005}
L. Kozlowski, G. Rossi, L. Blanquart, R. Marchesini, Y. Guang, G. Chow, J. Richardson, D. Standley, \emph{Pixel noise suppression via SoC management of tapered reset in a 1920x1080 CMOS image sensor}, \emph{IEEE Sensors Journal} {\bf 40} (2005) 2766-2775.

\bibitem{spieler2005}
H. Spieler, \emph{Semiconductor Detector Systems}, in \emph{Semiconductor Science and Technology} \textbf{12} Oxford University Press, Oxford U.K. (2005).

\bibitem{koerner2010}
L. J. Koerner.
\emph{X-ray analog pixel array detector for single synchrotron bunch time-resolved imaging.}, 
\emph{Ph.D. dissertation} (2010) Cornell University, USA. \url{https://ecommons.cornell.edu/handle/1813/17584}

\bibitem{wegmann1987}
G. Wegmann, E.A. Vittoz, F. Rahali, \emph{Charge Injection in Analog MOS Switches}, \emph{IEEE Journal of Solid-State Circuits} {\bf 22} (1987) 1091-1097.

\bibitem{becker2018}
J. Becker, M.W. Tate, K.S. Shanks, H.T. Philipp, J.T. Weissa, P. Purohit, D. Chamberlain and S.M. Gruner.
\emph{Characterization of chromium compensated GaAs as an X-ray sensor material for charge-integrating pixel array detectors}, 
\emph{JINST} {\bf 13} (2018) P01007.

\bibitem{shanks2019}
J. Shanks, J. Barley, S. Barrett, M. Billing, G. Codner, Y. Li, X. Liu, A. Lyndaker, D. Rice, N. Rider, D.L. Rubin, A. Temnykh, S.T. Wang
\emph{Accelerator design for the Cornell High Energy Synchrotron Source upgrade},
\emph{Phys. Rev. Accel. Beams} {\bf 22} (2019) 021602.

\bibitem{temnykh2015}
A. B. Temnykh, A. Lyndaker, M. Kokole, T. Milharcic, J. Pockar, R. Geometrante 
\emph{Construction of CHESS compact undulator magnets at Kyma},
\emph{SPIE Optics+Optoelectronics. Advances in X-ray Free-Electron Lasers Instrumentation III} {\bf 9512} (2015) 05

\bibitem{temnykh2013}
A. Temnykh, D. Dale, E. Fontes, Y. Li, A. Lyndaker, P. Revesz, D. Rice and A. Woll
\emph{Compact Undulator for the Cornell High Energy Synchrotron Source: Design and Beam Test Results}, \emph{J. Phys.: Conf. Ser.} {\bf 425} (2013) 032004.


\bibitem{trueb2012}
P. Trueb, B.A. Sobott, R. Schnyder, T. Loeliger, M. Schneebeli, M. Kobas, R.P. Rassool, D.J. Peake, C. Broennimann \emph{Improved count rate corrections for highest data quality with PILATUS detectors}, \emph{J. Synchrotron Rad.} {\bf 19} (2012) 347-351.

% Please avoid comments such as "For a review'', "For some examples",
% "and references therein" or move them in the text. In general,
% please leave only references in the bibliography and move all
% accessory text in footnotes.

% Also, please have only one work for each \bibitem.


\end{thebibliography}
\end{document}